\documentclass[letter,11pt]{article}
\usepackage{etex}
\usepackage[top=1in,bottom=1in,left=1in,right=1in]{geometry}
\usepackage{slashed}
\usepackage{epsfig}
\usepackage{comment}
\usepackage{cancel}
\usepackage{bbm}
\usepackage{array}
\usepackage{bigints}
\usepackage{booktabs}
\usepackage{color}
\usepackage{dsfont}
\usepackage{float}
\usepackage{framed}
\usepackage{graphicx}
\usepackage{indentfirst}
\usepackage{mathrsfs}
\usepackage{multirow} 
\usepackage{subdepth}
\usepackage{subfig}
\usepackage{titlesec}
\usepackage[dotinlabels]{titletoc}
\usepackage{wrapfig}
\usepackage[all]{xy}
\usepackage{young}
\usepackage[vcentermath]{youngtab}
\usepackage{relsize}
\usepackage{hyperref}
\usepackage{feynmp}
\usepackage{appendix}
\numberwithin{equation}{section}

\usepackage[utf8]{inputenc}
\usepackage{geometry}
\usepackage{graphicx}
\usepackage{array}
\usepackage{subfig}
\usepackage{slashed}
\usepackage{amsmath}
\usepackage{amsfonts}
\usepackage{amssymb}
\usepackage{cite}
\usepackage{epsfig}
\usepackage{comment}
\usepackage{tikz}
\usepackage{feynmp}
\usepackage{cancel}
\usepackage{mathrsfs}
\usepackage{mathtools}

\usepackage{cleveref}
\crefname{section}{§}{§§}
\Crefname{section}{§}{§§}

 \def\p{\partial}

\def\0{{(0)}}
\def\1{{(1)}}
\def\2{{(2)}}
 
\def\co{{\cal O}}

\def\ci{{\mathscr I}}

\def\<{\langle }
\def\>{\rangle }

\newcommand{\bea}{\begin{eqnarray}}
\newcommand{\eea}{\end{eqnarray}}
\def\be{\begin{equation}}
\def\ee{\end{equation}}
\newcommand{\ba}{\begin{align}}
\newcommand{\ea}{\end{align}}

\renewcommand{\epsilon}{\varepsilon}

   \makeatletter
  \let\over=\@@over \let\overwithdelims=\@@overwithdelims
  \let\atop=\@@atop \let\atopwithdelims=\@@atopwithdelims
  \let\above=\@@above \let\abovewithdelims=\@@abovewithdelims
\renewcommand\section{\@startsection {section}{1}{\z@}%
                                   {-3.5ex \@plus -1ex \@minus -.2ex}
                                   {2.3ex \@plus.2ex}%
                                   {\normalfont\large\bfseries}}

\renewcommand\subsection{\@startsection{subsection}{2}{\z@}%
                                     {-3.25ex\@plus -1ex \@minus -.2ex}%
                                     {1.5ex \@plus .2ex}%
                                     {\normalfont\bfseries}}

\newcommand{\beq}{\begin{equation}}
\newcommand{\eeq}{\end{equation}}
\newcommand{\beqa}{\begin{eqnarray}}
\newcommand{\eeqa}{\end{eqnarray}}
\newcommand{\beqar}{\begin{eqnarray*}}

\newcommand{\ve}{{\varepsilon}}

\def\[{\big[}
\def\]{\big]}

\def\ve{{\varepsilon}}

\def\o{{\omega}}
\def\a{{\alpha}}
\def\b{{\beta}}

\def\bh{\bar{h}}

\def\CI{{\mathcal I}}
\def\ci{{\mathcal I}}

\def\CO{{\mathcal O}}

\def\+{{(+)}}
\def\-{{(-)}}
\def\0{{(0)}}
\def\1{{(1)}}
\def\2{{(2)}}
\def\3{{(3)}}

\newcommand{\bd}[1]{\begin{fmffile}{#1}\begin{fmfgraph*}}
\newcommand{\ed}{\end{fmfgraph*}\end{fmffile}}

\begin{document}
\begin{titlepage}
\unitlength = 1mm
\ \\
\vskip 3cm
\begin{center}

{\LARGE{\textsc{Gravitational Memory in Higher Dimensions}}}

\vspace{0.8cm}
Monica Pate, Ana-Maria Raclariu and Andrew Strominger

\vspace{1cm}

{\it  Center for the Fundamental Laws of Nature, Harvard University,\\
Cambridge, MA 02138, USA}

\vspace{0.8cm}

\begin{abstract}

It is shown that there is a universal gravitational memory effect measurable by inertial detectors in even spacetime dimensions $d\geq 4$. The effect  falls off at large radius $r$ as $r^{3-d}$. Moreover this memory effect sits at one corner of an 
infrared triangle with the other two corners occupied by Weinberg's soft graviton theorem and 
infinite-dimensional asymptotic symmetries. 

 \end{abstract}

\vspace{1.0cm}

\end{center}

\end{titlepage}

\pagestyle{empty}
\pagestyle{plain}

\def\vx{{\vec x}}
\def\p{\partial}
\def\po{$\cal P_O$}

\pagenumbering{arabic}
 

\tableofcontents

\section{Introduction}
In  four spacetime dimensions, a triangular equivalence has been established between Weinberg's soft graviton theorem, a subgroup of past and future BMS symmetries, and the gravitational memory effect \cite{He:2014laa,Strominger:2014pwa,Strominger:2017zoo,Strominger:2013jfa}, indicating  a rich universality in the deep infrared (IR). 
Many manifestations of this universal IR triangle have been found in a variety of systems. 
In this paper we address gravitating systems and construct an IR triangle in even dimensions greater than four. 

A puzzle immediately arises. In a number of papers, it has been claimed that there is no 
gravitational memory or BMS symmetry above $d=4$ \cite{Hollands:2003ie,Hollands:2003xp,Tanabe:2009va,Tanabe:2010rm,Tanabe:2011es,Tanabe:2012fg, Hollands:2016oma, Garfinkle:2017fre}.  On the other hand, the soft graviton theorem holds in any number of dimensions, and the associated infinity of conservation laws/symmetries should follow from matching conditions near spatial infinity. The full triangle including symmetries and memories should be traceable starting from the soft theorem corner. Indeed the memory effect is just a Fourier transform with respect to time of the soft theorem. 

In this paper we resolve this puzzle. Metric components contain both radiative and Coulombic terms which fall off 
like ${ r^{1-{d\over 2}}}$ and $ r^{3-d}$ respectively. These fall offs are the same only in $d=4$.
We will find a memory effect for any even dimension $d$ which falls off as $r^{3-d}$ like the Coulombic components. This does not in any way contradict the results of \cite{Hollands:2016oma, Garfinkle:2017fre} which consider memory effects only at order ${ r^{1-{d\over 2}}}$. Moreover we show that the memory effect exhibits universal behavior, meaning that for any even $d$, it sits at a corner of a triangle that includes both the soft graviton theorem and asymptotic symmetries/conservation laws.

A number of interesting phenomena arise along the way. In particular, the Goldstone boson of the associated symmetry breaking on the boundary of $\ci$ is hidden in a subradiative component of the metric appearing at the order of  Coulombic terms in the form of an undetermined integration constant in the perturbative  large-$r$ solution.  The leading soft graviton theorem in higher dimensions is not related to IR divergences and has a structure which resembles subleading soft theorems in $d=4$ \cite{Kapec:2014opa, Campiglia:2014yka}, for which this analysis may contain some lessons.

Overlapping results were independently obtained by Mao and Ouyang in \cite{Mao:2017wvx} and Campiglia and Coito in \cite{Campiglia:2017xkp}.

This paper is organized as follows. In section 2, we present relevant formulas of linearized gravity in harmonic gauge, impose boundary conditions on the metric perturbations and determine the resulting residual gauge symmetry. 
In section 3, we focus on gravity in six dimensions for which we show that the residual gauge symmetry found in the previous section generates a gravitational memory effect, thereby establishing these residual diffeomorphisms as physical asymptotic symmetries.
In section 4, we show that Weinberg's soft theorem in $d = 6$ computes the shift in the transverse metric component associated with the gravitational memory effect. Moreover, we reinterpret the soft theorem as a conservation law resulting from a matching condition of a component of the Weyl tensor at $\CI_+^-$ and $\CI_-^+$. In section 5, we generalize our analysis from $d = 6$ to all higher even-dimensional spacetimes.

 \section{General relativity in $d=2m+2>4$}

This section presents some basic formulas for asymptotically flat spacetimes in  $d=2m+2$ dimensions for $m\ge 2$\footnote{A similar harmonic gauge analysis is possible in $d=4, m=1$, but a separate analysis is needed as some of the equations degenerate.}  and integral. 

\subsection{Linearized gravity in harmonic gauge}

Consider perturbations $g_{\mu\nu} = g_{\mu\nu}^{(0)} + h_{\mu\nu}$ around a flat background in $d = 2 + 2m$ dimensions, $m \geq 2$. Here $g_{\mu\nu}^{(0)}$ denotes the higher-dimensional flat metric in retarded coordinates,
\begin{equation} \label{Bondicoord495}
ds^2 =g_{\mu\nu}^{(0)} dx^\mu dx^\nu = -du^2 - 2du dr + r^2\gamma_{AB} dz^A dz^B
\end{equation}
where $z^A$, $A = 1, \cdots, 2m$ are coordinates on the asymptotic $S^{2m}$.
Defining the trace-reversed perturbation
\begin{equation}
\bar{h}_{\mu\nu} = h_{\mu\nu} - \frac{1}{2}g_{\mu\nu}^{(0)} h, \qquad h = g^{(0)\mu\nu} h_{\mu\nu},
\end{equation}
and imposing the harmonic gauge condition
\begin{equation} \label{gaugecond045}
\nabla^{\mu} \bar{h}_{\mu\nu} = 0,
\end{equation}
the linearized Einstein equation becomes
\begin{equation} \label{einsteineq43}
\Box \bar{h}_{\mu\nu} = -16\pi G T_{\mu\nu}, \qquad  \Box \bar{h}_{\mu\nu} \equiv \nabla^{\rho}\nabla_{\rho} \bar{h}_{\mu\nu} .
\end{equation}
In an asymptotic analysis near null infinity we may include in $T_{\mu\nu}$ all forms of radiative stress-energy including gravity waves. In components
\begin{equation}
\label{lee}
\begin{split}
\Box \bar{h}_{uu} &= \left(\partial_r^2 - 2\partial_r\partial_u - \frac{2m}{r}\left(\partial_u - \partial_r \right) + r^{-2}D^2 \right)\bar{h}_{uu}, \\
\Box \bar{h}_{ur} &=
(\partial_r^2 - 2\partial_r\partial_u + r^{-2}D^2)\bar{h}_{ur} + \frac{2m}{r^2}(\bar{h}_{uu} - \bar{h}_{ur}) - \frac{2m}{r}(\partial_u - \partial_r)\bar{h}_{ur} - \frac{2}{r^3}D^A\bar{h}_{uA} ,\\ 
\Box \bar{h}_{rr} &= 
\left(\partial_r^2 - 2\partial_u\partial_r + r^{-2}D^2 \right)\bar{h}_{rr} - \frac{4}{r^3}D^A\bar{h}_{Ar} - \frac{2m}{r}\left(\partial_u \bar{h}_{rr} - \partial_r \bar{h}_{rr} \right) + \frac{4m}{r^2}\left(\bar{h}_{ur} - \bar{h}_{rr}\right) + \frac{2}{r^4}\gamma^{CB} \bar{h}_{CB},\\
\Box \bar{h}_{uA} &= 
\left(\partial_r^2 - 2\partial_u\partial_r + r^{-2}D^2 \right) \bar{h}_{uA} - \frac{2m - 2}{r}(\partial_u - \partial_r) \bar{h}_{uA} - \frac{2}{r}\partial_A\left(\bar{h}_{uu} - \bar{h}_{ur} \right) + \frac{1 - 2m}{r^2}\bar{h}_{uA},\\
\Box \bar{h}_{rA}  &= 
\left(\partial_r^2 - 2\partial_u\partial_r + r^{-2}D^2 \right)\bar{h}_{rA} - \frac{2m - 2}{r}\left(\partial_u - \partial_r \right)\bar{h}_{rA} - \frac{2}{r^3}D^C\bar{h}_{CA} - \frac{2}{r}\partial_A(\bar{h}_{ru} - \bar{h}_{rr})  \\
& \quad \quad+ \frac{2m + 2}{r^2}(\bar{h}_{Au} - \bar{h}_{Ar}) + \frac{1}{r^2}\bar{h}_{rA} - \frac{2m}{r^2}\bar{h}_{rA},\\
\Box \bar{h}_{AB} &= 
  \left(\partial_r^2 - 2\partial_u\partial_r + r^{-2}D^2 \right)\bar{h}_{AB} - \frac{2}{r}D_A\left(\bar{h}_{uB} - \bar{h}_{rB} \right) - \frac{2}{r}D_B\left(\bar{h}_{uA} - \bar{h}_{rA} \right)  \\
  & \quad \quad +\frac{4 - 4m}{r^2}\bar{h}_{AB} - \frac{4 - 2m}{r}\partial_r\bar{h}_{AB} + \frac{4 - 2m}{r}\partial_u \bar{h}_{AB} + 2\gamma_{AB}\left(\bar{h}_{uu} - 2\bar{h}_{ru} + \bar{h}_{rr} \right),
\end{split}
\end{equation}
where here and hereafter, $D_A$ denotes the covariant derivative with respect to the unit round metric $\gamma_{AB}$ on $S^{2m}$ and $D^2 = \gamma^{AB} D_A D_B$.
Likewise, the components of the harmonic gauge condition \eqref{gaugecond045}  are 
\beq\label{hgc}
				\begin{split}	
					\nabla^\mu \bh_{\mu u} & = - \p_u \bh_{ur} - \p_r \bh_{uu} + \p_r \bh_{ur}- \frac{2m}{r} (\bh_{uu}- \bh_{ur})+ \frac{1}{r^2}D^A \bh_{uA},\\
					\nabla^\mu \bh_{\mu r} & = - \p_u \bh_{rr} - \p_r \bh_{ur} + \p_r \bh_{rr}- \frac{2m}{r} (\bh_{ur}- \bh_{rr})+ \frac{1}{r^2}D^A \bh_{rA}- \frac{1}{r^3} \gamma^{AB} \bh_{AB},\\
					\nabla^\mu \bh_{\mu A} & = - \p_u \bh_{rA} - \p_r \bh_{uA} + \p_r \bh_{rA}- \frac{2m}{r} (\bh_{uA}- \bh_{rA})+ \frac{1}{r^2}D^B \bh_{BA} .
				\end{split}
\eeq
The residual diffeomorphisms $\xi^\mu$ that preserve the harmonic gauge condition \eqref{gaugecond045} obey $\Box \xi^{\mu} = 0$, or equivalently 
\begin{equation}
\label{rgt}
\begin{split}
\left(\partial_r^2 - 2\partial_r\partial_u + r^{-2}D^2 \right)\xi_u - \frac{2m}{r}(\partial_u \xi_u - \partial_r \xi_u) = 0,\\
(\partial_r^2 - 2\partial_r\partial_u + r^{-2}D^2)\xi_r - \frac{2}{r^3}D^A \xi_A - \frac{2m}{r}\left(\partial_u - \partial_r \right)\xi_r +\frac{2m}{r^2}\left(\xi_u - \xi_r \right) = 0,\\
(\partial_r^2 - 2\partial_r\partial_u + r^{-2}D^2)\xi_A - \frac{2m - 2}{r}(\partial_u - \partial_r) \xi_A + \frac{1 - 2m}{r^2}\xi_A - \frac{2}{r}\partial_A(\xi_u - \xi_r) = 0.  
\end{split}
\end{equation}

\subsection{Boundary conditions and solution space}

In $2m+2$ dimensions, radiative solutions of the wave equation fall off like ${1 \over r^m}$ in a local orthonormal frame, while Coulombic solutions have the faster (for $m>1$) falloff ${1 \over r^{2m-1}}$. We accordingly adopt the boundary conditions 
\begin{equation}
\label{bc}
\begin{split}
h_{uu} \sim \CO({r^{-2m+1}}),\qquad \qquad h_{ru} \sim \CO(r^{-2m}), \qquad \qquad h_{rr} \sim \CO(r^{-m-2}),\\
h_{uA}\sim \CO(r^{-2m+1}),  \qquad \qquad h_{rA} \sim \CO(r^{-m}), \qquad  \qquad h_{AB} \sim  \CO(r^{-m+2}),\\
h = g^{(0)}{}^{\mu\nu} h_{\mu\nu} \sim \CO(r^{-2m}), \qquad \qquad \gamma^{AB} h_{AB} \sim  \CO(r^{-m}).
\end{split}
\end{equation}
Here the $ h_{rr}$, $ h_{rA}$ and $h_{AB}$ components comprise the radiative modes and appear accordingly at $\co(r^{-m})$ in an orthonormal frame, while the Coulombic modes include
$h_{uu}$, $h_{uA}$, and $h_{ur}$ and appear at $\co(r^{-2m+1})$ in an orthonormal frame.\footnote{Strictly speaking, these components are not all independent meaning that 
 the Einstein equation together with the harmonic gauge condition require some of the components to fall off faster than a radiative or Coulombic mode in an orthonormal frame.}
Making  this division requires application of the residual gauge symmetry \eqref{rgt}, details of which are spelled out in Appendix A. 
Moreover, the metric components are not all independent but are related by the constraints \eqref{constraints321} as well as the harmonic gauge condition \eqref{gaugecond045}.
We  further impose the consistent fall-off conditions on the Einstein tensor
\beq
\label{etf}
		\begin{split}
			G_{uu} \sim \co (r^{-2m}), \quad \quad \quad G_{ur} \sim \co (r^{-(2m+2)}), \quad \quad \quad G_{rr} \sim \co (r^{-(2m+2)}), \\
			G_{uA} \sim \co (r^{-2m}), \quad \quad \quad G_{rA} \sim \co (r^{-(2m+1)}), \quad \quad \quad G_{AB} \sim \co (r^{-(2m-1)}),
		\end{split}
\eeq 
and assume that the components of the energy-momentum tensor $T_{\mu\nu}$ fall off at the same rate as  $G_{\mu\nu}$.  These boundary conditions allow the higher-dimensional generalization of Kerr \cite{Myers:1986un} as well as gravity waves. 
 Allowing for the difference in gauge choice, they are  consistent with the falloffs employed in \cite{Hollands:2016oma}, but are stronger than those  in \cite{Kapec:2015vwa}.  

In the large-$r$ limit, we assume an asymptotic expansion in inverse powers of $r$ of the metric perturbations, starting at the the order given by \eqref{bc}
\begin{equation}
\label{exp}
	\begin{split}
		h_{uu} = \sum_{n = 2m-1}^{\infty} \frac{h_{uu}^{(n)}}{r^n}, \qquad  h_{ur} = \sum_{n = 2m}^{\infty} \frac{h_{ur}^{(n)}}{r^n}, \quad h_{rr} = \sum_{n = m+2}^{\infty} \frac{h_{rr}^{(n)}}{r^n},\\
		h_{uA} = \sum_{n = 2m-1}^{\infty} \frac{h_{uA}^{(n)}}{r^n}, \qquad  h_{rA} = \sum_{n = m}^{\infty} \frac{h_{rA}^{(n)}}{r^n}, \quad h_{AB} = \sum_{n = m-2}^{\infty} \frac{h_{AB}^{(n)}}{r^n}.
	\end{split}
\end{equation}
We use these expansions to solve \eqref{gaugecond045} and \eqref{einsteineq43} order-by-order in $r$ and find that the coefficients of the expansions \eqref{exp} are all determined in terms of the radiative data $h_{AB}^{(m-2)}$, up to potentially significant $u$-independent integration constants. 
In particular, the Einstein equation \eqref{einsteineq43} together with the gauge condition \eqref{gaugecond045} imply the following constraints when $m \leq n \leq 2m-1$
\beq \label{constraints321} 
				\begin{split} 
					0=& h_{rr}^{(n)} + \gamma^{AB} h_{AB}^{(n-2)},\\
					0=&  \left(D^2-(n-2 m)^2-n\right)h_{rr}^{(n)}  +2  (m-n) \gamma^{AB} h_{AB}^{(n-2)}  -2  (m-n+1) D^A h_{rA}^{(n-1)} ,   \\
					 0=& \left(D^2-(2 m-n-1) (2 m-n)+2 (n-1)\right)D^A h_{rA}^{(n-1)}  +2  (n-m)D^A D^B h_{AB}^{(n-2)}+2 D^2 h_{rr}^{(n)}.
				\end{split}
			\eeq 
Using the above constraints together with the Einstein equation, one obtains a relation between the radiative mode and components of the metric  appearing at $\co(r^{-2m+1})$ 
\begin{equation}
\label{rmrec}
\begin{split}
\p_u^{m-1} D^A D^B h_{AB}^{(2m-3)} = \frac{(-1)^{m-1}}{2^{m-1} (m-1)!}\frac{D^2(D^2 + 2) + 8 - 8m}{(D^2 - 2(2m-3))(D^2 - (2m-2))}\\
\prod_{\ell = m+1}^{2m-1} \left(D^2 - (2m-\ell)(\ell-1) \right)D^A D^B h_{AB}^{(m-2)},
\end{split}
\end{equation}
where inverse powers of $D^2$ denote Green's functions.

Finally, the relation between the components of the metric and the flux of energy and momentum which appears at leading order in the $uu$ component of the Einstein equation is simply 
\beq \label{Tuueq}
	-16 \pi G T_{uu}^{(2m)}  = 2(m-1)  \p_u h_{uu}^{(2m-1)}  .
\eeq

\subsection{Residual symmetries}

The harmonic gauge condition \eqref{gaugecond045} and the falloffs \eqref{bc} do not fully fix all of the 
gauge symmetry. Residual symmetries remain of the form 
  \begin{equation} \label{alloweddiff}
\xi_f = \frac{(m-2) D^2f}{r^{2m-2}}\p_u - \frac{D^2 + 2(m-1)(2m-5)}{2(2m-1)}\frac{D^2 f}{r^{2m-2}}\p_r + \frac{D^A((D^2-4) m+4)}{ (2m-1)  }\frac{f}{r^{2m-1}}\p_A,
\end{equation}   
where $f$ is any function on the sphere. Poincar\'e  transformations are also allowed diffeomorphisms but are peripheral to our discussion and so
are not explicitly included  in \eqref{alloweddiff}.
Under this residual symmetry, $h_{AB}^{(2m-3)}$ transforms  as 
\begin{equation} \label{habshift}
	\begin{split}
		\delta_{\xi_f} h_{AB}^{(2m-3)} & = \frac{1}{2m-1} \left(2D_AD_B\left((D^2-4) m+4\right)-\gamma_{AB}D^2 \left(D^2 + 2(m-1)(2m-5)\right) \right) f,\\
		D^A D^B\delta_{\xi_f}  h_{AB}^{(2m-3)} &= (8 + D^2 (2 + D^2) - 8 m)D^2f,
	\end{split}
\end{equation}
while  for $m\ge 2$, the radiative mode $h^{(m-2)}_{AB}$ is unaffected.
\section{Gravitational memory in $d=6$}  
\label{sgm}
In this section we derive the higher-dimensional gravitational memory effect. For notational simplicity, we focus on the $d = 6$ case which illustrates many of the features of the most general case. The generalization to all even higher dimensions
 is discussed in section \ref{s7}.

We begin by considering two inertial detectors near $\mathcal{I}^+$ moving along timelike geodesics\footnote{There will be corrections to the tangent vectors in \eqref{tangvect} subleading in $r$  but these will not affect our analysis.} with tangent vector
\begin{equation} \label{tangvect}
k = \p_u, \qquad k_\mu k^\mu = -1.
\end{equation}
The relative transverse displacement $s^A$ of the detectors obeys
\begin{equation}
\label{gde}
\frac{d^2 s^A}{du^2} = R^A{}_{ uu B}s^B.
\end{equation}
Using the metric perturbation boundary conditions \eqref{bc}, the linearized Riemann tensor is
\begin{equation}
R^A{}_{uuB} = \frac{\gamma^{AC}}{2r^2}\p_u^2 h_{CB}+ \co(r^{-4}). 
\end{equation}  

Let us consider a scenario in which the system is in vacuum ($i.e.$ stationary) at initial and final times, $u_i$ and $u_f$ respectively, while during the interval $u_i < u< u_f$ there is a transit of gravitational radiation.   
To leading order in the large-$r$ expansion, \eqref{gde} can be integrated directly to find
\begin{equation}
\Delta s^A = \frac{\gamma^{AC}}{2r^2}\Delta h_{CB} s^B_i+ \co(r^{-4}),
\end{equation}
where $s_i^A = s^A(u_i)$ is the separation of the detectors at some initial retarded time $u = u_i$, $\Delta s^A = s^A(u_f) - s^A(u_i),$ and $ \Delta h_{AB} = h_{AB}(u_f) - h_{AB}(u_i)$.  Expanding at large $r$, 
\begin{equation} \Delta h_{AB}=  \frac{\Delta h_{AB}^{(1)}}{r}+\cdots . \end{equation}
The would-be leading term vanishes because 
\be\label{sw} \Delta h_{AB}^{(0)}=0.\ee
One way to derive this is to note that the initial and final configurations are radiative vacua, but according to 
\eqref{habshift}, one can always choose $h_{AB}^{(0)}=0$ in the vacuum. 
Therefore the leading term is `frozen' and cannot be shifted by the passage of waves.  This conclusion was arrived at by a different method in \cite{Hollands:2016oma}, and is the basis of the statement that there is no gravitational memory in higher dimensions. However it really only implies that the gravitational memory effect does not appear at the radiative order of the large-$r$ expansion, rather it appears at the Coulombic order which is subleading above $d=4$.

Using \eqref{habshift} to write $\Delta h_{AB}^{(1)}$ in terms of pure gauge configurations, one finds 
\begin{equation}\label{tih}
\Delta s^A = \frac{\gamma^{AC}}{2r^3}\left[ \frac{4}{3}\left(D_B D_C - \frac{1}{4}\gamma_{BC}D^2\right)(D^2 -2) \Delta C\right] s^B_i + \mathcal{O}(r^{-4}),
\end{equation}
where $C$ is a component on the sphere of a vacuum metric which obeys 
\be \delta_{\xi_f}C=f \ee and accordingly shall be referred to as the Goldstone mode.  $C$ characterizes the vacuum configuration at a given retarded time and  $\Delta C \equiv C|_{u_f} - C|_{u_i}$.

Thus, we conclude that the large diffeomorphism  \eqref{alloweddiff} which distinguishes the late and early vacua can be measured by gravitational memory experiments which are sensitive to the ${1 \over r^3}$ Coulombic components of the metric near null infinity. Since they can be measured, \eqref{alloweddiff} are not trivial diffeomorphisms and occupy one corner of the IR  triangle.  The relation \eqref{tih} provides the side of the triangle which connects these large diffeomorphisms ($i.e.$ asymptotic symmetries) to the memory effect. 
We now proceed to complete the triangle with the third corner.

\section{Weinberg's soft theorem in $d=6$}

\label{sec:softthm}
	Continuing in $d=6$, in this section we show that Weinberg's soft graviton theorem is a formula for the 
	shift in precisely the same metric component $\Delta h^{(1)}_{AB}.$ One way to demonstrate this is to Fourier transform the usual momentum space formulas for the soft theorem.  This relates the zero mode  of the radiative piece
	(${1 \over r^2}$ in an orthonormal frame) of the metric $h^{(0)}_{AB}$, which in turn is related to the shift in  $h^{(1)}_{AB}$ via the $u$ integral of \eqref{rmrec} (or equation \eqref{intmem} below) to the classical radiation field sourced by the scattering process.  This Fourier relation was worked out in \cite{Mao:2017wvx}. Here we shall proceed via a different route, showing that the soft formula can be expressed as a conservation law following from antipodal matching conditions at null infinity,  as in $d=4$. This conservation law then readily yields  a formula for the memory shift $\Delta h^{(1)}_{AB}$. 
\subsection{Soft theorem as Ward identity}

We begin as in  \cite{Kapec:2015vwa} by rewriting the soft theorem as a Ward identity. Consider fluctuations of the $d=6$ metric about a flat background, $g_{\mu\nu} = \eta_{\mu\nu} + \kappa \tilde h_{\mu\nu}$, where $\kappa^2 = 32\pi G$.\footnote{  The `$~\tilde{}~$'  indicates that these 
graviton modes are normalized differently from the ones appearing in the rest of the paper.}
The radiative degrees of freedom of the gravitational field have the mode expansion
\begin{equation}
\label{mode-exp}
\tilde h_{\mu\nu} = \sum_{\alpha} \int \frac{d^5q}{(2\pi)^5}\frac{1}{2\omega_q}\left[\varepsilon_{\mu\nu}^{*\alpha}a_{\alpha}(\vec{q})e^{iq\cdot x} + \varepsilon_{\mu\nu}^{\alpha} a_{\alpha}^{\dagger}(\vec{q}) e^{-iq\cdot x} \right],
\end{equation}
where $\omega_q = |\vec{q}|$ and $\varepsilon_{\mu\nu}^{\alpha}(\vec{q})$ are polarization tensors obeying
\begin{equation}
\label{pt}
q^{\mu}\varepsilon^{\alpha}_{\mu\nu}(\vec{q}) = 0, \qquad  2 \sum_\alpha \ve_{\alpha}^*{}^{ij}(\vec q) \ve_{\alpha}^{k \ell}(\vec q) = \pi^{ik} \pi^{j \ell} +  \pi^{i \ell} \pi^{j k} - \frac{1}{2} \pi^{ij} \pi^{k \ell}, \qquad \pi^{ij} = \delta^{ij}- \frac{q^i q^j}{\vec q^2}  .
\end{equation} 
The modes $a_{\a}$ and $a^{\dagger}_{\a}$ obey the canonical commutation relations 
\be 
[a_{\a}(\vec{p}), a_{\b}^{\dagger}(\vec{q})] = 2\omega_{q}\delta_{\a\b}(2\pi)^5\delta^5(\vec{p} - \vec{q}).
\ee
Note  \eqref{pt}  directly implies that the mode expansion \eqref{mode-exp} obeys the harmonic gauge condition.  The utility of this gauge choice in the context of the IR triangle was originally suggested in \cite{Avery:2015gxa}. Also see \cite{Campiglia:2015kxa, Campiglia:2016jdj,Campiglia:2016efb} for additional analyses employing harmonic gauge.

The free radiative data at $\mathcal{I}^+$ is
\begin{equation}
h_{AB}^{(0)}(u, z) \equiv \kappa \lim_{r\rightarrow \infty} \p_A x^{\mu} \p_B x^{\nu} \tilde h_{\mu\nu}(u + r, r\hat{x}(z))  
\end{equation}
and can be evaluated by taking  a saddle point approximation at large $r$  of \eqref{mode-exp}. The result is  \cite{Kapec:2015vwa}
\begin{equation}
h_{AB}^{(0)} = -\frac{2\pi^2\kappa}{(2\pi)^5}\p_A\hat{x}^i\p_B\hat{x}^j\sum_{\alpha}\int d\omega_q \omega_q\left[\varepsilon_{ij}^{*\alpha}a_{\alpha}(\omega_q\hat{x})e^{-i\omega_q u} + \varepsilon_{ij}^{\alpha} a_{\alpha}^{\dagger}(\omega_q\hat{x}) e^{i\omega_q u} \right].
\end{equation}
The frequency space expression is obtained by performing a Fourier transform 
\begin{equation}
h_{AB}^{ \omega (0)} = -\frac{\kappa}{2(2\pi)^2}\p_A\hat{x}^i\p_B\hat{x}^j\sum_{\alpha}\int d\omega_q \omega_q\left[\varepsilon_{ij}^{*\alpha}a_{\alpha}(\omega_q\hat{x}) \delta (\omega - \omega_q) 
	+ \varepsilon_{ij}^{\alpha} a_{\alpha}^{\dagger}(\omega_q\hat{x}) \delta (\omega + \omega_q)  \right].
\end{equation}
Hence,
\be
h_{AB}^{\o(0)} = -\frac{\kappa \o}{8\pi^2}\p_A\hat{x}^i\p_B\hat{x}^j\sum_{\a}\varepsilon_{ij}^{*\a}a_{\a}(\o\hat{x}), \qquad h_{AB}^{-\o(0)} = -\frac{\kappa \o}{8\pi^2}\p_A\hat{x}^i\p_B\hat{x}^j\sum_{\a}\varepsilon_{ij}^{\a}a_{\a}^{\dagger}(\o\hat{x}),
\ee
where $\o >0$.
The zero frequency mode of $h_{AB}$ is then defined in terms of a linear combination of positive and negative zero-frequency Fourier modes
\begin{equation}
\label{zm}
\int du h_{AB}^{(0)} = \frac{1}{2}\lim_{\omega\rightarrow 0}\left(h_{AB}^{-\omega (0)} + h_{AB}^{\omega (0)} \right).
\end{equation} 

The soft graviton theorem for an outgoing soft graviton can be expressed in terms of an insertion of \eqref{zm} into the $n\to m$ particle  $\mathcal{S}$-matrix
\begin{equation}
\label{st6}
\langle z_{n+1},...|\int du h_{AB}^{(0)}(u,z)\mathcal{S}| z_1,...\rangle = -\frac{\kappa^2}{2(4\pi)^2}F_{AB}^{\text{out}}(z;z_k)\langle z_{n+1},...|\mathcal{S}| z_1,...\rangle, 
\end{equation}
where we have labelled the asymptotic particles by their four-momenta  $p^\mu_k = E_k \left(1, \hat x(z_k)\right)$ and \begin{equation}
\label{sf6}
F_{AB}^{\text{out}}(z;z_k) \equiv \omega \p_A\hat{x}^i\p_B\hat{x}^j\sum_{\alpha}\varepsilon_{ij}^{*\alpha}\left[\sum_{k = n+1}^{n+m}\frac{\varepsilon_{\mu\nu}^{\alpha} p^{\mu}_kp^{\nu}_k}{p_k\cdot q} - \sum_{k=1}^n \frac{\varepsilon_{\mu\nu}^{\alpha} p^{\mu}_k p^{\nu}_k}{p_k\cdot q}\right].
\end{equation} 
When acted upon by the following differential operator, \eqref{sf6}  localizes to a sum over $\delta$-functions
\beq
	\sqrt{\gamma} \left(D^2-2\right) D^A D^B F_{AB}^{\text{out}}(z;z_k) = 3 (4 \pi)^2\left[\sum_{k = n+1}^{n+m} E_k \delta^4(z-z_k) - \sum_{k=1}^n E_k \delta^4(z-z_k)\right].
\eeq
Then acting with this differential operator on \eqref{st6} and convolving  against an arbitrary function on the sphere $D^2 f(z)$, one finds 
\begin{equation}
\label{soft}
\begin{split}
-\frac{1}{3\kappa^2}\int d^4z \sqrt{\gamma}D^2 f (z)&(D^2 - 2 )D^A D^B \langle z_{n+1}, ... | \int du h_{AB}^{(0)}(z)\mathcal{S}|z_1,...\rangle\\
& =  \frac{1}{2}\left[\sum_{k = n+1}^{n+m} E_k D^2f (z_k) - \sum_{k=1}^n E_k D^2 f (z_k)\right]\langle z_{n+1},...|\mathcal{S}|z_1,...\rangle.
\end{split}
\end{equation}
Combined with the result of an analogous analysis near $\mathcal{I}^-$, \eqref{soft} leads to the Ward identity   
\begin{equation}\label{stm}
\langle z_{n+1},...|Q^+ \mathcal{S} - \mathcal{S}Q^-| z_1,...\rangle = 0,
\end{equation}
where the charges can be decomposed into hard and soft pieces
\begin{equation}
Q^{\pm} = Q_H^{\pm} + Q_S^{\pm}.
\end{equation}
The action of the hard charges on outgoing asymptotic states is reproduced by the  action of the leading order $uu$ component of the energy-momentum  
\begin{equation} \label{qh6}
\langle z_{n+1},...|Q_H^+ = \langle z_{n+1},...|\sum_{k = n+1}^{n+m} E_k D^2 f (z_k) = \langle z_{n+1},...|\int_{\ci^+} dud^4z \sqrt{\gamma}D^2 f  T_{uu}^{(4)},
\end{equation}
while the soft charges,  defined as
\begin{equation} \label{qs6}
Q_S^+ = \frac{1}{3\kappa^2}\int du d^4z \sqrt{\gamma}D^2 f (z)(D^2 - 2)D^A D^B h_{AB}^{(0)},
\end{equation}
add soft gravitons to the asymptotic state.  Similar expressions can be determined for charges $Q^-$ which act  near $\mathcal{I}^-$.

We now show that $Q_S^+$ is proportional to the shift  $\Delta C $ which appears in the memory formula \eqref{tih}.
The $AB$ component of the linearized Einstein equations in $d=6$ at $\mathcal{O}(r^{-2})$ (or equivalently equation \eqref{rmrec}) is 
\begin{equation}
(D^2 - 4)h_{AB}^{(0)} + 2\p_u h_{AB}^{(1)} = 0,
\end{equation}
which implies
\begin{equation} \label{intmem}
\int du D^A D^B h_{AB}^{(0)}   =- \frac{2}{D^2 + 4}  D^A D^B  \Delta h_{AB}^{(1)} .
\end{equation} 
Recall that  $\Delta h_{AB}^{(1)}$ is pure gauge 
\beq
	 D^A D^B  \Delta h_{AB}^{(1)}  = (D^2-2 )(D^2+4)D^2\Delta C,
\eeq
where $\Delta C$ is the shift in the Goldstone mode across ${\cal I}^+$.  It follows that 
\be \label{qsC} Q_S^+= -\frac{2}{3\kappa^2}\int d^4z \sqrt{\gamma} D^2 f (z)(D^2 - 2)^2 D^2 \Delta C. \ee

Substituting these expressions in  \eqref{soft}, one finds that the soft theorem computes the expectation value of angular  derivatives of $ \Delta C$ in the presence of a scattering event 
\begin{equation}
\label{mem}
\begin{split}
\frac{2}{3\kappa^2}&\int  d^4z \sqrt{\gamma}D^2f (z) (D^2 - 2)^2 D^2 \langle z_{n+1}, ... | \Delta C(z)\mathcal{S}|z_1,...\rangle \\
 &= \frac{1}{2}\left[\sum_{k = n+1}^{n+m} E_kD^2 f (z_k) - \sum_{k=1}^n E_kD^2f (z_k)\right]\langle z_{n+1},...|\mathcal{S}|z_1,...\rangle.
\end{split}
\end{equation}
As anticipated in section \ref{sgm}, we find that gravitational radiation emitted during generic scattering processes shifts the component of the metric $h_{AB}^{(1)}$ where the precise shift 
is given by   the gauge transformation law \eqref{habshift} with $f=\Delta C$ as given in  \eqref{mem}.

\subsection{Conservation law}

\label{conlaw6}

The soft theorem in the form \eqref{stm} equates the  outgoing integral $Q^+$ on ${\cal I}^+$ to an incoming  integral $Q^-$ on ${\cal I}^-$.   In this subsection we show that both of these are given by $S^4$  boundary integrals near spatial infinity, and their equality is implied by the antipodal matching conditions. Moreover the conserved charges are arbitrary moments of the Weyl tensor component whose zero mode is the Bondi mass.  

	To see this explicitly, recall the charge $Q^+$ given by its decomposition into  hard  \eqref{qh6} and soft  \eqref{qs6} parts
	\beq
		\begin{split}
			Q^+  = \int_{\ci^+} du d^4z \sqrt{\gamma} D^2f  \left(T_{uu}^{(4)}- \frac{2}{3 \kappa^2} \frac{D^2 - 2}{D^2+4} D^A D^B \p_u h_{AB}^{(1)}\right).
		\end{split}
	\eeq
Using \eqref{Tuueq} and performing the $u$ integral, one finds
	\beq 
		\begin{split}
			Q^+  =\frac{1}{16 \pi G} \int_{S^4}  d^4z \sqrt{\gamma} D^2f  \left(-2  h_{uu}^{(3)}- \frac{1}{3 } \frac{D^2 - 2}{D^2+4} D^A D^B   h_{AB}^{(1)}\right) \Bigg|^{\ci^+_+}_{\ci^+_-}.
		\end{split}
	\eeq
This combination of gauge fields is proportional to the leading contribution to the asymptotic expansion of the following component of the Weyl tensor evaluated in a radiative vacuum near the boundaries of $\cal I$ 
	\beq \label{cruru6d}
		C_{ruru} = \frac{1}{r^5}\left(- \frac{1}{2}\frac{(D^2-2)   }{ (D^2+4) } D^A D^B   h_{AB}^{(1)}-3   h_{uu}^{(3)}  \right) + \co \big(r^{-6}\big).
	\eeq
	Note the expression is non-local in the gauge field only because we chose to express it in terms of $h_{AB}^{(1)}$.\footnote{Equivalent local expressions can be obtained for example by substituting $h_{AB}^{(1)}$ for $h_{rA}^{(2)}$.}
	
	Natural boundary conditions on the Weyl tensor manifestly lead to  sufficiently fast fall off  behavior of the $C_{ruru}$ component at $\ci^+_+$ for the charges to be written as local expressions at $\ci^+_-$\footnote{We have not verified that this charge in any sense generates via the Dirac bracket the large diffeomorphisms of section 2.3. Such a demonstration would have to carefully account for all our subsidiary gauge fixing as well as the mismatching powers of $r$ found in Appendix C.}
	\beq \label{Weylcharge}
		\begin{split}
		Q^+ & = \frac{1}{16 \pi G} \int_{\ci^+_-}  d^4z \sqrt{\gamma} D^2f \left(2  h_{uu}^{(3)}+ \frac{1}{3 } \frac{D^2 - 2}{D^2+4} D^A D^B   h_{AB}^{(1)}\right) \\
			&=-\frac{2}{3} \frac{1}{16 \pi G} \lim_{r \to \infty} \int_{\ci^+_-}  d^4z \sqrt{\gamma}r^5 D^2f  C_{ruru} .
		\end{split}
	\eeq
	In particular, if we replace $D^2f$ by $1$, then we recover the total mass  
	\be \left. Q^+\right|_{D^2f \to 1}= \int_{\ci^+} du d^4z \sqrt{\gamma}  \left(T_{uu}^{(4)}- \frac{2}{3 \kappa^2} \frac{D^2 - 2}{D^2+4} D^A D^B \p_u h_{AB}^{(1)}\right), \ee
	where notice that the second term is a total derivative on $S^4$ and thus vanishes upon integration.
Antipodal matching of this  Weyl tensor component on ${\ci^-_+}$ to that on ${\ci^+_-}$ then implies directly 
\be Q^+=Q^-.\ee

In summary there is a complete IR triangle in $d=6$ with the leading soft graviton theorem in one corner,  ${1 \over r^3}$ gravitational memory effect at the second corner and asymptotic symmetries/conservation laws at the third. 
\section{Higher dimensional generalization}	
\label{s7}

	In this section, we generalize the analysis in sections \ref{sgm} and \ref{sec:softthm} to all even dimensions $d>4$.

\subsection{Geodesic deviation in higher dimensions}
\label{gdalld}

	In general dimensions, the relative transverse displacement $s^A$ of the detectors due to the transit of gravitational radiation is given by \eqref{gde}. 
To linear order in the metric perturbations
\begin{equation}
R^A{}_{uuB} = \frac{\gamma^{AC}}{2r^2}\left(\p_u^2 h_{CB} - D_B\p_u h_{uC} - D_C\p_u h_{uB} + D_BD_C h_{uu} \right) - \frac{\delta^A_B}{2r}\left(2\p_u h_{ru} - \p_r h_{uu} - \p_u h_{uu} \right).
\end{equation}
Using the fall-off conditions \eqref{bc}, one finds this component is given by an  asymptotic expansion of the form
\beq
	R^A{}_{uuB}  =  \frac{\gamma^{AC}}{2r^2}\sum_{k = m-2}^{2m-3} \frac{\p_u^2 h_{CB}^{(k)}}{r^k}+ \mathcal{O}(r^{-2m}).
\eeq
Integrating \eqref{gde} twice, one finds
\begin{equation}
\label{hdgde}
\Delta s^A =  \gamma^{AC} \frac{\Delta h_{CB}^{(2m-3)}}{2r^{2m-1}} s_i^B + \mathcal{O}(r^{-2m}),
\end{equation}
where just like in the six-dimensional case, we assume the system is in vacuum at initial and final times.  This means  that the modes $ h_{CB}^{(n)}$ up to $n = 2m-3$ do not contribute since 
their  vacuum configurations do not undergo a relative displacement between early and late times or equivalently, they do not change under \eqref{alloweddiff}.

Finally, the displacement $\Delta s^A$ can be expressed directly in terms of the change in the vacuum configuration
\beq
	\Delta s^A  =  \frac{   \left(2D^AD_B\left((D^2-4) m+4\right)-\delta^A_B D^2 \left(D^2 + 2(m-1)(2m-5)\right) \right) \Delta C}{2(2m-1)r^{2m-1}} 
		 s_i^B + \mathcal{O}(r^{-2m}),
\eeq
where $C$ is a function on $S^{2m}$ that parameterizes the space of inequivalent vacuum configurations.

 We conclude that gravitational memory experiments sensitive to the $\frac{1}{r^{2m-1}}$ Coulombic components of the metric near null infinity can measure the shift between early and late vacua generated by \eqref{alloweddiff}.

\subsection{Soft theorem as Ward identity}

From a stationary phase approximation of the standard mode expansion in higher dimensions,  we find  
\begin{equation}
\label{rmhd}
\begin{split}
h_{AB}^{(m-2)}(u,z) &= \frac{\kappa \p_A\hat{x}^i\p_B\hat{x}^j}{4\pi (2\pi i)^m}\sum_{\alpha} \int d\omega_q \omega_q^{m-1}\left[\ve_{ij}^{*\alpha} a_{\alpha}(\omega_q\hat{x}) e^{-i\omega_q u} + (-1)^m\ve_{ij}^{\alpha} a_{\alpha}^{\dagger}(\omega_q \hat{x}) e^{i\omega_q u} \right]  \\
&= \p_u^{m-1}~ \frac{i^{m-1}\kappa \p_A\hat{x}^i\p_B\hat{x}^j}{4\pi (2\pi i)^m}\sum_{\alpha}\int d\omega_q \left[\ve_{ij}^{*\alpha} a_{\alpha}(\omega_q\hat{x}) e^{-i\omega_q u} - \ve_{ij}^{\alpha} a_{\alpha}^{\dagger}(\omega_q \hat{x}) e^{i\omega_q u} \right].
\end{split}
\end{equation}
The finite frequency modes are given by the Fourier transform 
\beq
h_{AB}^{ \omega (m-2)} = \frac{(-i)^m\kappa}{2(2\pi)^m}\p_A\hat{x}^i\p_B\hat{x}^j\sum_{\alpha}\int d\omega_q \omega_q^{m-1}\left[\varepsilon_{ij}^{*\alpha}a_{\alpha}(\omega_q\hat{x}) \delta (\omega - \omega_q) 
	+(-1)^m \varepsilon_{ij}^{\alpha} a_{\alpha}^{\dagger}(\omega_q\hat{x}) \delta (\omega + \omega_q)  \right],
\eeq
or equivalently
\be
	\begin{split}
h_{AB}^{\o(m-2)}& = \frac{(-i)^m\kappa}{2(2\pi)^m}\p_A\hat{x}^i\p_B \hat{x}^j \o^{m-1}\sum_\alpha \varepsilon_{ij}^{*\a} a_{\a}(\o\hat{x}),\\
  h_{AB}^{-\o(m-2)}& = \frac{i^m\kappa}{2(2\pi)^m}\p_A\hat{x}^i\p_B \hat{x}^j \o^{m-1}\sum_\alpha \varepsilon_{ij}^{\a} a_{\a}^{\dagger}(\o\hat{x}),
	\end{split}
\ee
with $\o >0$.
The soft theorem can be written as
\begin{equation}
\label{std}
\langle z_{n+1},...|h_{AB}^{0(m-2)}(z)\mathcal{S}| z_1,...\rangle = -\frac{(-1)^m\kappa^2}{8(2\pi)^m}F_{AB}^{\text{out}}(z;z_k)\langle z_{n+1},...|\mathcal{S}| z_1,...\rangle, 
\end{equation}
with $F_{AB}^{\text{out}}(z;z_k)$ given in \eqref{sf6} where $A,B$ now run over the coordinates of $S^{2m}$ and the zero mode $h_{AB}^{0(m-2)}$ is defined as
\begin{equation}
\label{zmd}
h_{AB}^{0(m-2)} = \frac{1}{2}\lim_{\omega\rightarrow 0}(i\omega)^{2-m}\left(h_{AB}^{\omega (m-2)} + (-1)^m h_{AB}^{-\omega (m-2)} \right).
\end{equation}

 Although we just obtained an expression for the soft graviton mode starting from the mode expansion of the radiative component of the metric, as in six dimensions, the zero frequency graviton 
 is more naturally related to the shift in a Coulombic component of the metric.  In particular, by comparing the mode expansion \eqref{rmhd} with the expression  \eqref{rmrec},
 we obtain an expression for the frequency mode expansion of the component of the metric associated to the memory
\begin{equation}
\label{memory-me}
\begin{split}
&\frac{2^{m-1}(m-1)!}{(-1)^{m-1}}H(u,z) = \\
&\left(\prod_{\ell= m+1}^{2m-1} \mathcal{D}_\ell\right) D^A D^B \left[\frac{\kappa \p_A \hat{x}^i \p_B\hat{x}^j}{4\pi i (2\pi)^m} \sum_{\alpha}\int d\omega_q\left(\ve_{ij}^{*\alpha} a_{\alpha}(\omega_q \hat{x}) e^{-i\omega_q u} - \ve_{ij}^{\alpha} a^{\dagger}_{\alpha}(\omega_q \hat{x}) e^{i\omega_q u} \right) \right],
\end{split}
\end{equation}
where
\begin{equation}
\mathcal{D}_\ell = D^2 - (2m - \ell)(\ell-1)
\end{equation}
and $H(u,z)$ contains the memory metric component
\begin{equation} \label{defhuz}
H(u,z) = \frac{(D^2 - 2(2m-3))(D^2 - (2m-2))}{D^2(D^2 + 2) + 8 - 8m} D^A D^Bh_{AB}^{(2m-3)}.
\end{equation}
Equation \eqref{memory-me} can now be inverted to express the creation and annihilation operators in terms of $H(u,z)$. Extracting the zero-mode, we find that
\begin{equation}
\begin{split}
& \int du \langle z_{n+1},...| \p_u H(u,z)\mathcal{S}|z_1,...\rangle = \\
&\lim_{\omega\rightarrow 0} \frac{(-1)^m}{2^{m-1} (m-1)!}\left(\prod_{\ell = m+1}^{2m-1} \mathcal{D}_\ell\right) D^A D^B\left[ \frac{\kappa \p_A \hat{x}^i \p_B\hat{x}^j}{8 \pi (2\pi)^m} \sum_{\alpha}2\pi \varepsilon_{ij}^{*\alpha}\omega \langle z_{n+1},...| a_{\alpha}(\omega\hat{x})\mathcal{S}|z_1,... \rangle\right].
\end{split}
\end{equation}

Using the action of the differential operators $\mathcal{D}_\ell$ on soft factor $F_{AB}$ derived in \cite{Kapec:2015vwa}, the soft theorem \eqref{std} can equivalently be written as follows
\begin{equation}
\label{hd-memory}
	\begin{split}
 & (D^2 - 2(2m-3))(D^2 - (2m-2)) D^2 \langle z_{n+1},...| \Delta C \mathcal{S}|z_1,...\rangle \\&
= \frac{\kappa^2 (2m - 1)}{4 \sqrt{\gamma}}\left[\sum_{k = n+1}^{n+m} E_k \delta^{2m}(z-z_k) - \sum_{k=1}^n E_k \delta^{2m}(z-z_k) \right]\langle z_{n+1},...| \mathcal{S}|z_1,...\rangle,
	\end{split}
\end{equation}
where we have used \eqref{habshift} to express the difference between the early and late vacua across $\CI^+$ in terms of the shift $\Delta C$ in the Goldstone mode across $\CI^+$. We note that the form of the large diffeomorphism  \eqref{alloweddiff}  ensures the Green's function appearing in  \eqref{defhuz} cancels when the change in $h_{AB}^{(2m-3)}$ across $\mathcal{I}^+$ is pure gauge.

Integrating \eqref{hd-memory} against an arbitrary function $D^2 f$ on $S^{2m}$, we find
\begin{equation}
\label{hd-st}
	\begin{split}
	 & \int d^{2m}z \sqrt{\gamma} D^2 f(z) (D^2 - 2(2m-3))(D^2 - (2m-2)) D^2 \langle z_{n+1},...| \Delta C \mathcal{S}|z_1,...\rangle \\&
= \frac{(2m - 1)\kappa^2}{4} \left[\sum_{k = n+1}^{n+m} E_k D^2 f(z_k) - \sum_{k=1}^n E_k D^2 f(z_k) \right]\langle z_{n+1},...| \mathcal{S}|z_1,...\rangle.
	\end{split}
\end{equation}

The higher dimensional analog of the soft charge \eqref{qsC} is then
\begin{equation}
Q_S^+ = -\frac{2}{(2m - 1)\kappa^2}\int d^{2m} z\sqrt{\gamma} D^2 f(z)(D^2 - 2(2m-3))(D^2 - (2m-2)) D^2 \Delta C .
\end{equation}
Combined with the corresponding formula for $\mathcal{I}^-$ as well as the higher dimensional generalizations of the hard charges \eqref{qh6}, we deduce that the Ward identity 
\begin{equation}
\langle z_{n+1},...|Q^+\mathcal{S} - \mathcal{S}Q^-| z_1,...\rangle = 0
\end{equation}
computes the shift in $h_{AB}^{(2m-3)}$ along $\mathcal{I}$ induced by a flux of gravitational radiation. This shift is given by the gauge transformation \eqref{habshift} with $f = \Delta C$ as in \eqref{hd-st}.

\subsection{Conservation law}

\label{sec:conslaw}

In this subsection, we generalize the conclusions of section \ref{conlaw6} to arbitrary higher even dimensions. We first show that the conserved charges can be expressed in terms of arbitrary moments of the Weyl tensor component whose zero mode is the Bondi mass. From \eqref{hd-st} we find that the full charge $Q^+$ takes the following form
\be 
Q^+ = \int_{\CI^+} du d^{2m}z \sqrt{\gamma} D^2 f \left(T_{uu}^{(2m)} - \frac{2}{(2m-1)\kappa^2}\frac{(D^2 - 2(m-1))(D^2 - 2(2m-3))}{D^2(D^2 + 2) - 8m + 8}\p_u D^A D^B h_{AB}^{(2m-3)} \right).
\ee

Using \eqref{Tuueq} and performing the $u$ integral, this becomes
\be
\begin{split}
 Q^+ = &\frac{1}{16\pi G}\int_{S^{2m}} d^{2m}z \sqrt{\gamma} D^2 f\\&\left. \left(-2(m-1)h_{uu}^{(2m-1)} - 
 \frac{1}{2m-1}\frac{(D^2 - 2(m-1))(D^2 - 
 2(2m-3))}{D^2(D^2 + 2) - 8m + 8} D^A D^B h_{AB}^{(2m-3)} \right)\right|_{\CI^+_-}^{\CI^+_+}.
 \end{split}
\ee

In direct analogy with the discussion of section \ref{conlaw6}, we find that the leading term in the $1/r$ expansion of the $C_{ruru}$ component of the Weyl tensor evaluated in a radiative vacuum near the boundary of $\ci^+$  is
\begin{equation}
C_{ruru}^{(2m+1)} =  - (2m - 1)(m-1)h_{uu}^{(2m-1)} - \frac{1}{2}\frac{(D^2 - 2(m-1))(D^2 - 2(2m-3))}{D^2(D^2 + 2) - 8m + 8}D^A D^Bh_{AB}^{(2m-3)}.
\end{equation} 
Assuming that $C_{ruru}$ vanishes sufficiently fast at $\CI^+_+$, the total charge on $\CI^+$ can be written as
\be  \label{alldQ}
Q^+ = -\frac{2}{(2m-1) 16\pi G}\int_{\CI^+_-}  d^{2m}z \sqrt{\gamma} D^2f   C_{ruru}^{(2m+1)}.
\ee
 
The antipodal matching condition of the Weyl tensor at $\ci^+_-$ and $\ci^-_+$ then naturally leads to the interpretation of the memory effect as the direct consequence of conservation law 
\be
Q^+ = Q^-
\ee
along $\ci$.

In conclusion, this shows that the soft theorem in higher dimensions is a consequence of the symmetry associated with the diffeomorphism \eqref{alloweddiff}. Moreover, \eqref{alloweddiff} generates the higher-dimensional analog of the $4d$ gravitational memory effect. It is a `large' gauge transformation in the sense that it computes the leading shift in the metric due to gravitational flux which is measured by the geodesic deviation equation discussed in section \ref{gdalld}. This completes the triangular equivalence between soft theorems, asymptotic symmetries and gravitational memory in higher even-dimensional asymptotically flat spacetimes.

\section*{Acknowledgements}
We are grateful to Temple He and Dan Kapec for useful conversations and especially to Sasha Zhiboedov for collaboration during the early stages of this project. This work was supported by NSF grant 1205550 and the John Templeton Foundation.

 \begin{appendices}

	\section{Asymptotic expansions}
		Assuming a $1/r$ asymptotic expansion of the metric perturbation and using the notation $F^{(n)}$ to denote the coefficient of  $r^{-n}$, the Einstein equation components are
		\beq \label{einsteineq948}
				\begin{split}
					\left[\square \bh_{uu}\right]^{(n)}&  = 2(n-m-1)  \p_u \bh_{uu}^{(n-1)} + \left[D^2-(n-2)(2m-n+1)\right]  \bh_{uu}^{(n-2)} ,\\
					\left[\square \bh_{ur}\right]^{(n)} & =  2(n-m-1)\p_u \bh_{ur}^{(n-1)} + \left[D^2-(n-1)(2m-n+2)  \right]  \bh_{ur}^{(n-2)} +  2m \bh_{uu}^{(n-2)} -   2 D^A \bh_{uA}^{(n-3)},  \\
					\left[\square \bh_{rr}\right]^{(n)} & = 2(n-m-1) \p_u \bh_{rr}^{(n-1)} + \left[D^2 - (n-2) (2 m - n+1) \right]  \bh_{rr}^{(n-2)} +  4m  \left (\bh_{ru}^{(n-2)} - \bh_{rr}^{(n-2)}\right)\\& \quad 
													   - 4  D^A \bh_{rA}^{(n-3)}  +  2  \gamma^{AB} \bh_{AB}^{(n-4)} , \\ 
					\left[\square \bh_{uA}\right]^{(n)} & =  2(n-m) \p_u \bh_{uA}^{(n-1)} + \left[D^2-(n-1)(2m-n)-1  \right] \bh_{uA}^{(n-2)}-  2   \p_A\bh_{uu}^{(n-1)} +2 \p_A\bh_{ur}^{(n-1)} ,\\
					\left[\square \bh_{rA}\right]^{(n)} & = 2(n-m) \p_u \bh_{rA}^{(n-1)} + \left[D^2-n(2m-n+1)-3  \right] \bh_{rA}^{(n-2)}- 2 \p_A  \bh_{ur}^{(n-1)} +2 \p_A \bh_{rr}^{(n-1)}  \\  
										& \quad - 2  D^B \bh_{AB}^{(n-3)} +(2m+2) \bh_{uA}^{(n-2 )},\\
					\left[\square \bh_{AB}\right]^{(n)} & =   2(n-m+1)\p_u \bh_{AB}^{(n-1)} +  \left[D^2-2   + n - 2 m n + n^2 \right] \bh_{AB}^{(n-2)} \\& \quad
										-2\left(D_A \bh_{uB}^{(n-1)} - D_A \bh_{rB}^{(n-1)} + D_B \bh_{uA}^{(n-1)} - D_B \bh_{rA}^{(n-1)}  \right)+ 2 \gamma_{AB} \left(\bh_{uu}^{(n)} - 2 \bh_{ur}^{(n)}+ \bh_{rr}^{(n)}\right) ,
				\end{split}
			\eeq
			while the components of the harmonic gauge condition are
			\beq \label{gaugeconst495} 
				\begin{split}	
					\left[\nabla^\mu \bh_{\mu u}\right]^{(n)} & = - \p_u \bh_{ur}^{(n)} - (2m-n+1)(\bh_{uu}^{(n-1)}- \bh_{ur}^{(n-1)})+ D^A \bh_{uA}^{(n-2)} ,\\
					\left[\nabla^\mu \bh_{\mu r}\right]^{(n)} & = - \p_u \bh_{rr}^{(n)} - (2m-n+1)(\bh_{ur}^{(n-1)}- \bh_{rr}^{(n-1)})+  D^A \bh_{rA}^{(n-2)} - \gamma^{AB} \bh_{AB} ^{(n-3)} ,\\
					\left[\nabla^\mu \bh_{\mu A}\right]^{(n)} & = - \p_u \bh_{rA}^{(n)} -(2m-n+1) (\bh_{uA}^{(n-1)}- \bh_{rA}^{(n-1)})+  D^B \bh_{BA}^{(n-2)} .
				\end{split}
			\eeq
			The expansion of the harmonic gauge condition on residual diffeomorphisms  \eqref{rgt}  is
			\beq   \label{harmonicdiff958}
				\begin{split}
					\left[\square \xi_u\right]^{(n)} &=   2 (n-m-1 ) \p_u \xi_u^{(n-1)} + \left[ D^2- (1 + 2 m - n) (n-2) \right]  \xi_u^{(n-2)}  ,  \\
					\left[\square \xi_r\right]^{(n)} &=  2 (n-m-1 ) \p_u \xi_r^{(n-1)} +  \left [  D^2- (2 + 2 m - n) (n-1) \right]  \xi_r ^{(n-2)}+  2m  \xi_u^{(n-2)} -  2  D^A \xi_A^{(n-3)}  , \\
					\left[\square \xi_A\right]^{(n)} &=  2 (n-m) \p_u \xi_A^{(n-1)} +  \left [ D^2 + (-1 + n) (-2 m + n) - 1 \right] \xi_A^{(n-2)} - 2 \p_A (\xi_u^{(n-1)} - \xi_r^{(n-1)}) . 
				\end{split}
			\eeq
			
			\section{Residual gauge-fixing}
				\label{appgaugefix} 
		
		In this appendix, we provide the details of a consistent gauge-fixing procedure by which we arrive at the boundary conditions for the metric perturbation employed in this paper \eqref{bc}.
	
		We start with the following weak boundary conditions on the metric 
	\begin{equation}
\label{bc1r}
\begin{split}
\bar{h}_{uu} \sim \CO({r^{-m}}),\qquad \qquad \bar{h}_{ru} \sim \CO(r^{-m}), \qquad \qquad\bar{h}_{rr} \sim \CO(r^{-m}),\\
\bar{h}_{uA}\sim \CO(r^{-m+1}),  \qquad \qquad \bar{h}_{rA} \sim \CO(r^{-m+1}), \qquad  \qquad\bar{h}_{AB} \sim  \CO(r^{-m+2}).
\end{split}
\end{equation}
		These were selected such that the leading order terms are allowed (by the Einstein equation) to be free functions of $(u, z^A)$.  In particular, 
		they can be determined by looking for the order at which the coefficient of the $\p_u \bh_{\mu\nu}$ term in \eqref{einsteineq948} vanishes.  If one works with weaker boundary conditions than these, then 
				a consistent asymptotic expansion must include logarithmic terms in $r$.  Finally, note these imply $g^{\mu\nu} h_{\mu\nu} \sim \co(r^{-m})$.
				
		Combining the gauge constraints \eqref{gaugeconst495} with the components of the Einstein equation \eqref{einsteineq948} at orders in the asymptotic expansion when the Einstein tensor vanishes, we 
				obtain the following constraints  
			\beq \label{constraints405} 
				\begin{split}
					0=&(D^2-(2 m-n-1) (2 m-n))\bh_{ur}^{(n)} -2 (2 m-n-1) (n-m) \bh_{uu}^{(n)} +2 (n-m)  D^A \bh_{uA}^{(n-1)},  \\
					0=&  \left(D^2-(n-2 m)^2-n\right)\bh_{rr}^{(n)}+2 (m-(2 m-n-1) (n-m))\bh_{ur}^{(n)} +2  (m-n) \gamma^{AB} \bh_{AB}^{(n-2)} \\
					  			& \quad\quad\quad -2  (m-n+1) D^A \bh_{rA}^{(n-1)} ,   \\
					 0=& (D^2-(2 m-n-1) (2 m-n)+2 (n-1))D^A \bh_{rA}^{(n-1)}+2 (m-n) (2 m-n)D^A \bh_{uA}^{(n-1)}\\
					  	& \quad\quad\quad +2  (n-m)D^A D^B \bh_{AB}^{(n-2)}+2 D^2 (\bh_{rr}^{(n)}-\bh_{ur}^{(n)}),
				\end{split}
			\eeq
			where all three  apply when $m \leq n \leq 2m-1$.
			
			Evaluated at leading order $n=m$ these constraints imply
				\beq
					 \bh_{ur}^{(m)} = \bh_{rA}^{(m-1)} = \bh_{rr}^{(m)} = 0.
				\eeq
				
			The residual gauge transformations that preserve the boundary conditions\footnote{In particular, residual diffeomorphisms with weaker fall off conditions are ruled out by the fact that a consistent asymptotic expansion for 
			such diffeomorphisms must include logarithmic terms in $r$, which would in turn necessitate the addition of such terms in the asymptotic expansion of the metric.} \eqref{bc1r} are given to leading order by  
				\beq
					\xi = \frac{1}{r^m}\xi^u{}^{(m)}\p_u + \frac{1}{r^m} \xi^r{}^{(m)} \p_r+\frac{1}{r^{m+1}}\xi^A{}^{(m+1)}\p_A + \cdots,
				\eeq
				where $\xi^u{}^{(m)}$, $\xi^r{}^{(m)}$, and $\xi^A{}^{(m+1)}$ are free functions of $(u, z^A)$.   
			Under  these diffeomorphisms, the leading order fields transform in the following way 
			\beq \label{alldmetricdifftransf059}
				\begin{split} 
					\delta \bh_{uu}^{(m)}  = 2 \p_u \xi_u^{(m)}- \p_u \xi_r^{(m)}, \quad \quad \quad
						\delta \bh_{uA}^{(m-1)} = \p_u \xi_A^{(m-1)}, \quad \quad \quad 
							\delta h^{(m)} = -2 \p_u \xi_r^{(m)},\\
								\delta \bh_{AB}^{(m-2)} = \gamma_{AB}\p_u \xi_r^{(m)},  \quad \quad \quad  
						\delta \bh_{ur}^{(m)} =\delta \bh_{rr}^{(m)} =\delta \bh_{rA}^{(m-1)}  =0. 
				\end{split}
			\eeq
			
			We use the $u$-dependent part of the free functions $\xi^u{}^{(m)}$, $\xi^r{}^{(m)}$, and $\xi^A{}^{(m+1)}$ to set 
			\beq
				\bh_{uu}^{(m)} =  h^{(m)} =  \bh_{uA}^{(m-1)} = 0.
			\eeq
			Since $ \bh_{ur}^{(m)}  $ and $ \bh_{rr}^{(m)}  $ are zero, $ h^{(m)} = 0$ implies 
			\beq
				\gamma^{AB}\bh_{AB}^{(m-2)} = 0.
			\eeq
			Upon performing this gauge-fixing we are left with residual gauge transformations, which at leading order are given by $u$-independent $\xi^u{}^{(m)}$, $\xi^r{}^{(m)}$, and $\xi^A{}^{(m+1)}$.
			
			Now suppose one has gauge-fixed  
			\beq \label{gaugefixton059}
				\bh_{uu}^{(n')} =  h^{(n')} =  \bh_{uA}^{(n'-1)} = 0,  \quad \quad \quad n' \leq n.
			\eeq
			The remaining residual gauge transformations that preserve  \eqref{gaugefixton059} are  given to leading order by  
				\beq \label{residsymmatn}
					\xi = \frac{1}{r^n}\xi^u{}^{(n)}\p_u + \frac{1}{r^n} \xi^r{}^{(n)} \p_r+\frac{1}{r^{n+1}}\xi^A{}^{(n+1)}\p_A + \cdots,
				\eeq
				where $\xi^u{}^{(n)}$, $\xi^r{}^{(n)}$, and $\xi^A{}^{(n+1)}$ are $u$-independent free functions on the sphere $S^{2m}$.   Note with this gauge-fixing,
				 the  constraints \eqref{constraints405} imply that
				 \beq
				 	\bh_{ur}^{(n')} =0,  \quad \quad \quad n' \leq n.
				 \eeq
			Then, at this order, the following components of the Einstein equation significantly simplify
			\beq \label{simpleEE405}
				\begin{split}
					\p_u \bh_{uu}^{(n+1)} &= 0, \quad \quad \quad n \leq2m-3,\\
					\p_u \bh_{uA}^{(n)} &= 0, \quad \quad \quad n \leq 2m-2,\\
					\p_u h^{(n+1)}& = 0, \quad \quad \quad n \leq 2m-2,\\
				\end{split}
			\eeq
			implying that $ \bh_{uu}^{(n+1)} $, $  \bh_{uA}^{(n)}$ and $h^{(n+1)}$ are $u$-independent.
			
			The transformation of these components under the residual gauge symmetry \eqref{residsymmatn} is
			\beq \begin{split}
					\delta h^{(n+1)} &= -2 \p_u \xi_r^{(n+1)} +2(n-2m) (\xi_u^{(n)} -    \xi_r^{(n)}) + 2 D^A \xi_A^{(n-1)},   \\
					\delta \bh_{uu}^{(n+1)} & =  2 \p_u \xi_u^{(n+1)} - \p_u \xi_r^{(n+1)}  +(n- 2m) (\xi_u^{(n)} -\xi_r^{(n)}) +  D^A \xi_A^{(n-1)}  ,\\ 
					\delta \bh_{uA}^{(n)}   &= \p_u \xi_A^{(n)} + \p_A \xi_u^{(n)},
			\end{split} \eeq
			where using \eqref{harmonicdiff958}, $\p_u \xi_r^{(n+1)}$, $\p_u \xi_u^{(n+1)}$ and $\p_u \xi_A^{(n)}$ can be written in terms of $\xi_u^{(n)}$, $\xi_r^{(n)}$ and $\xi_A^{(n-1)} $.
			
			These transformations are linearly independent and thus can be used to set 
			\beq
				\bh_{uu}^{(n+1)} =  h^{(n+1)} =  \bh_{uA}^{(n)} = 0,
			\eeq
			leaving  the remaining residual gauge symmetry  with  leading order terms now given by $u$-independent $\xi^u{}^{(n+1)}$, $\xi^r{}^{(n+1)}$, and $\xi^A{}^{(n+2)}$.  
			
			Hence, we can iterate this process until $n$ surpasses the limits in \eqref{simpleEE405} and thus we can set
			\beq \label{totalgaugefix495}
			\begin{split}
				\bh_{uu}^{(n)}= \bh_{ur}^{(n)}= 0,     \quad \quad \quad m \leq n \leq 2m-2 , \\
				\bh_{uA}^{(n-1)}  =h^{(n)} =0,  \quad \quad \quad m \leq n \leq 2m-1 . 
			\end{split}
			\eeq 
			Upon performing this residual gauge-fixing, the first constraint in  \eqref{constraints405}  is identically zero for $n \leq 2m-2$ and sets $\bh_{ur}^{(2m-1)} = 0$ when $n = 2m-1$.    Moreover, 
			note the vanishing of the trace up to and including  order $r^{-2m+1}$ will allow us henceforth to work with the original metric perturbations $h_{\mu\nu}$ instead of the trace-reversed metric perturbations $\bh_{\mu\nu}$.

			 The vanishing trace condition together with the second two constraints in 
			 \eqref{constraints405}  simplify  for $ m \leq n \leq 2m-1$ to the three constraints appearing in  \eqref{constraints321}. 
		When $n \leq 2m-1$, we can use \eqref{constraints321} to solve for the other components in terms of $h_{rr}^{(n)}$ 
		\beq \label{allhrr}
			\begin{split}
				 \gamma^{AB} h_{AB}^{(n-2)}&=-h_{rr}^{(n)} ,  \\
				 D^Ah_{rA}^{(n-1)} & = \frac{ \left(D^2-(2m-n)(1 + 2m - n)\right)}{2 (m-n+1)}h_{rr}^{(n)},  \\
				D^AD^Bh_{AB}^{(n-2)}& =\frac{(D^2-(2 m-n-1) (2 m-n)+1)^2+4 (m-n) (2 m-n) (2 m-n+1)-1}{4 (m-n) (m-n+1)}h_{rr}^{(n)}.
			\end{split}
		\eeq
		Note,  this gauge fixing procedure sets 
		\beq
			h_{rr}^{(m)} = h_{rr}^{(m+1)} =0
		\eeq
		so  the last two equations hold provided $n>m+1$.
		
		Nonetheless, we can explicitly solve the Einstein equations at the leading orders and we find 
		\beq
			 D^A \p_uh_{rA}^{(m)}  = D^A D^B h_{AB}^{(m-2)}  , \quad \quad \quad  \p_u h_{rr}^{(m+2)} =  D^A h_{rA}^{(m)}.
		\eeq	
		Together these imply
		\beq \label{du2hrr056}
			 \p_u^2 h_{rr}^{(m+2)}= D^A D^B h_{AB}^{(m-2)}.
		\eeq
		Then, for $m+2 \leq n \leq 2m-1 $,  using \eqref{allhrr}, the asymptotic expansion of the $r$ component of the gauge constraint \eqref{gaugeconst495}  becomes
		\beq \label{rgaugeconst597}
				\begin{split}	 
					\p_u h_{rr}^{(n+1)}& =  \frac{D^2-(n-2) (2 m-n+1)}{2 (m-n+1)}h_{rr}^{(n)} .
				\end{split}
		\eeq 
		Taking multiple derivatives and using \eqref{du2hrr056}, this becomes 
		\beq \label{relatetoradiative058}
			\p_u^{n} h_{rr}^{(m+n)} = \prod_{\ell = m+1}^{m+n-2} \frac{D^2-(\ell-1) (2 m-\ell)}{2 (m-\ell)}D^A D^B h_{AB}^{(m-2)}.
		\eeq 
		The memory term lies at
			\beq \label{memoryhabhrr}
				D^A D^B h_{AB}^{(2m-3)} = \frac{(D^2 (D^2+2)-8 m+8)  }{4 (m-2) (m-1)}h_{rr}^{(2m-1)}.
			\eeq  
			Using \eqref{relatetoradiative058}, we find
			\beq \label{DDmemoryalld}
				\begin{split}
				\p_u^{m-1} D^A D^B h_{AB}^{(2m-3)}  
				&= \frac{(D^2 (D^2+2)-8 m+8)  }{(D^2-4 m+6) (D^2-2 m+2)} \prod_{\ell= m+1}^{2m-1} \frac{D^2-(\ell-1) (2 m-\ell)}{2 (m-\ell)}D^A D^B h_{AB}^{(m-2)}.
				\end{split}
			\eeq

		 \section{Canonical charges}
		 Using a canonical covariant formalism \cite{Wald:1999wa,Barnich:2011mi}, one can construct charges that are associated to the symmetries of the theory.  In gravity, the charge associated to a generic diffeomorphism $\xi$ is given by
	\beq \label{cancharge}
					Q^{+}_{\text{can}} = - \frac{1}{16 \pi G} \int_{\ci^+_-} d^{2m} z \sqrt{\gamma} r^{2m} F_{ru},
				\eeq
				where 
				\beq
					\begin{split}
					F_{\mu\nu} &= \frac{1}{2} \left(\nabla_\mu \xi_\nu - \nabla_\nu\xi_\mu \right)h+ \left (\nabla_\mu h^\lambda{}_\nu - \nabla_\nu h^\lambda{}_\mu\right)\xi_\lambda
					 + \left (\nabla_\lambda \xi_\mu h^\lambda{}_\nu - \nabla_\lambda \xi_\nu h^\lambda{}_\mu \right)
					\\& \quad \quad  
								- \left (\nabla_\lambda h^\lambda{}_\nu \xi_\mu -\nabla_\lambda h^\lambda{}_\mu \xi_\nu\right) - \left (\nabla_\mu h \xi_\nu - \nabla_\nu h \xi_\mu\right).
					\end{split}
				\eeq
	The relevant term for our analysis in retarded Bondi coordinates \eqref{Bondicoord495} is  
				\beq   \label{Frugen}
					\begin{split}
						F_{ru} &= \xi^u \left (\p_r \bh_{uu}-  \p_u \bh_{ur}\right) 
									+ \xi^r \left (\p_r \bh_{ur}-\p_u \bh_{rr} \right) 
								- \bh_{uu} \p_r \xi_r + \bh_{ur} \left [\p_r \xi_u - \p_u\xi_r  +\p_r \xi_r   \right]\\ & \quad
									+\bh_{rr}  \left ( \p_u \xi_u-\p_r \xi_u  \right)   
								+ \frac{1}{r^2} \left (\bh_{Au} D^A \xi_r - \bh_{Ar} D^A \xi_u\right)+ \xi^A \left[\left(\p_r - \frac{2}{r}\right)\bh_{Au} - \p_u \bh_{Ar}\right].
					\end{split}
				\eeq 
	In $d = 2m+2$ dimensions, $F_{ru}$ evaluated on the large diffeomorphism \eqref{alloweddiff} in the absence of radiation is given to leading order by
	 \beq \label{FruallD}
					\begin{split}
						 F_{ru} &=- \frac{ 2(m-2)D^2f }{r^{4m-2}} \left(2  (m-1)   h_{uu}^{(2m-1)}+\frac{ (D^2-4 m+6) (D^2-2 m+2) }{   (2 m-1) (D^2 (D^2+2)-8 m+8)  }	D^A D^B h_{AB}^{(2m-3)} \right)\\& \quad \quad
						  + \co(r^{-4m+1})
					\end{split}
	\eeq
	up to a total derivative term which vanishes in \eqref{cancharge} upon integration over $S^{2m}$.
	Then,  substituting this expression in  \eqref{cancharge}, one finds in the large-$r$ limit that the  canonical charge associated to the large diffeomorphism \eqref{alloweddiff} is proportional to the conserved charge derived from the soft theorem \eqref{alldQ}
\begin{equation}
Q_{\text{can}}^+  = \frac{2(m-2)}{r^{2m-2}} Q^+ .
\end{equation} 

The idea that components of the charge which fall off as some power of $1/r$ could generate non-trivial symmetries was put forward in \cite{Conde:2016rom, Mao:2017tey}. In particular, \cite{Conde:2016rom} showed that the $4d$ subleading soft graviton theorem can be recovered from a conservation law associated with a subleading component of the supertranslation charges in a large-$r$ expansion. We leave the interpretation of charges with a large-$r$ falloff as well as a possible relation between leading soft theorems in higher dimensions and subleading soft theorems in lower dimensions to future investigation.

\end{appendices}

\providecommand{\href}[2]{#2}\begingroup\raggedright


\begin{thebibliography}{10}

\bibitem{Strominger:2013jfa} 
  A.~Strominger,
  ``On BMS Invariance of Gravitational Scattering,''
  JHEP {\bf 1407}, 152 (2014)
  [arXiv:1312.2229 [hep-th]].

\bibitem{He:2014laa} 
  T.~He, V.~Lysov, P.~Mitra and A.~Strominger,
  ``BMS supertranslations and Weinberg’s soft graviton theorem,''
  JHEP {\bf 1505}, 151 (2015)
  [arXiv:1401.7026 [hep-th]].
  
  \bibitem{Strominger:2014pwa} 
  A.~Strominger and A.~Zhiboedov,
  ``Gravitational Memory, BMS Supertranslations and Soft Theorems,''
  JHEP {\bf 1601}, 086 (2016)
  [arXiv:1411.5745 [hep-th]].
  
  \bibitem{Strominger:2017zoo} 
  A.~Strominger,
  ``Lectures on the Infrared Structure of Gravity and Gauge Theory,''
  arXiv:1703.05448 [hep-th].
  

  
\bibitem{Hollands:2003ie} 
  S.~Hollands and A.~Ishibashi,
  ``Asymptotic flatness and Bondi energy in higher dimensional gravity,''
  J.\ Math.\ Phys.\  {\bf 46}, 022503 (2005)  [gr-qc/0304054].
  
\bibitem{Hollands:2003xp} 
  S.~Hollands and A.~Ishibashi,
  ``Asymptotic flatness at null infinity in higher dimensional gravity,''
  hep-th/0311178.
  
\bibitem{Tanabe:2009va} 
  K.~Tanabe, N.~Tanahashi and T.~Shiromizu,
  ``On asymptotic structure at null infinity in five dimensions,''
  J.\ Math.\ Phys.\  {\bf 51}, 062502 (2010)
  [arXiv:0909.0426 [gr-qc]].
  
\bibitem{Tanabe:2010rm} 
  K.~Tanabe, N.~Tanahashi and T.~Shiromizu,
  ``Angular momentum at null infinity in five dimensions,''
  J.\ Math.\ Phys.\  {\bf 52}, 032501 (2011)
  [arXiv:1010.1664 [gr-qc]].
  
\bibitem{Tanabe:2011es} 
  K.~Tanabe, S.~Kinoshita and T.~Shiromizu,
 ``Asymptotic flatness at null infinity in arbitrary dimensions,''
  Phys.\ Rev.\ D {\bf 84}, 044055 (2011)
  [arXiv:1104.0303 [gr-qc]].
  
\bibitem{Tanabe:2012fg} 
  K.~Tanabe, T.~Shiromizu and S.~Kinoshita,
 ``Angular momentum at null infinity in higher dimensions,''
  Phys.\ Rev.\ D {\bf 85}, 124058 (2012)
  [arXiv:1203.0452 [gr-qc]].
  

\bibitem{Hollands:2016oma} 
  S.~Hollands, A.~Ishibashi and R.~M.~Wald,
  ``BMS Supertranslations and Memory in Four and Higher Dimensions,''
  Class.\ Quant.\ Grav.\  {\bf 34}, no. 15, 155005 (2017)  [arXiv:1612.03290 [gr-qc]].
  
  \bibitem{Garfinkle:2017fre} 
  D.~Garfinkle, S.~Hollands, A.~Ishibashi, A.~Tolish and R.~M.~Wald,
  ``The Memory Effect for Particle Scattering in Even Spacetime Dimensions,''
  Class.\ Quant.\ Grav.\  {\bf 34}, no. 14, 145015 (2017)
  [arXiv:1702.00095 [gr-qc]].
  
\bibitem{Kapec:2014opa} 
  D.~Kapec, V.~Lysov, S.~Pasterski and A.~Strominger,
  ``Semiclassical Virasoro symmetry of the quantum gravity $ \mathcal{S}$-matrix,''
  JHEP {\bf 1408}, 058 (2014)
  [arXiv:1406.3312 [hep-th]].
  
  \bibitem{Campiglia:2014yka} 
  M.~Campiglia and A.~Laddha,
  ``Asymptotic symmetries and subleading soft graviton theorem,''
  Phys.\ Rev.\ D {\bf 90}, no. 12, 124028 (2014)
  [arXiv:1408.2228 [hep-th]].
  
\bibitem{Mao:2017wvx} 
  P.~Mao and H.~Ouyang,
  ``Note on soft theorems and memories in even dimensions,''  arXiv:1707.07118 [hep-th].
  
\bibitem{Campiglia:2017xkp} 
  M.~Campiglia and L.~Coito,
  ``Asymptotic charges from soft scalars in even dimensions,''
  arXiv:1711.05773 [hep-th].
  
\bibitem{Myers:1986un} 
  R.~C.~Myers and M.~J.~Perry,
  ``Black Holes in Higher Dimensional Space-Times,''
  Annals Phys.\  {\bf 172}, 304 (1986).
  
\bibitem{Kapec:2015vwa} 
  D.~Kapec, V.~Lysov, S.~Pasterski and A.~Strominger,
  ``Higher-Dimensional Supertranslations and Weinberg's Soft Graviton Theorem,''
  Annals of Mathematical Sciences and Applications, Volume 2 (2017),
  pp 69-94
  [arXiv:1502.07644 [gr-qc]].
  
  \bibitem{Avery:2015gxa} 
  S.~G.~Avery and B.~U.~W.~Schwab,
  ``Burg-Metzner-Sachs symmetry, string theory, and soft theorems,''
  Phys.\ Rev.\ D {\bf 93}, 026003 (2016)
  [arXiv:1506.05789 [hep-th]].
  
  \bibitem{Campiglia:2015kxa} 
  M.~Campiglia and A.~Laddha,
  ``Asymptotic symmetries of gravity and soft theorems for massive particles,''
  JHEP {\bf 1512}, 094 (2015)
  [arXiv:1509.01406 [hep-th]].
  
  \bibitem{Campiglia:2016jdj} 
  M.~Campiglia and A.~Laddha,
  ``Sub-subleading soft gravitons: New symmetries of quantum gravity?,''
  Phys.\ Lett.\ B {\bf 764}, 218 (2017)
  [arXiv:1605.09094 [gr-qc]].
  
\bibitem{Campiglia:2016efb} 
  M.~Campiglia and A.~Laddha,
  ``Sub-subleading soft gravitons and large diffeomorphisms,''
  JHEP {\bf 1701}, 036 (2017)
  [arXiv:1608.00685 [gr-qc]].
  
\bibitem{Wald:1999wa} 
  R.~M.~Wald and A.~Zoupas,
  ``A General definition of 'conserved quantities' in general relativity and other theories of gravity,''
  Phys.\ Rev.\ D {\bf 61}, 084027 (2000)
  [gr-qc/9911095].
  
\bibitem{Barnich:2011mi} 
  G.~Barnich and C.~Troessaert,
  ``BMS charge algebra,''
  JHEP {\bf 1112}, 105 (2011)
  [arXiv:1106.0213 [hep-th]].
  
  
  \bibitem{Conde:2016rom} 
  E.~Conde and P.~Mao,
  ``BMS Supertranslations and Not So Soft Gravitons,''
  JHEP {\bf 1705}, 060 (2017)
  [arXiv:1612.08294 [hep-th]].
  
 
  \bibitem{Mao:2017tey} 
  P.~Mao and J.~B.~Wu,
  ``Note on asymptotic symmetries and soft gluon theorems,''
  Phys.\ Rev.\ D {\bf 96}, no. 6, 065023 (2017)
  [arXiv:1704.05740 [hep-th]].
 

  
  \end{thebibliography}
\end{document}